\documentclass[11pt,a4paper]{article}
 \usepackage{jheppub}

\pdfoutput=1

\usepackage{subfigure}
\usepackage{multirow}

\newcommand{\eqn}[1]{eq.~(\ref{#1})}
\newcommand{\Eqn}[1]{Eq.~(\ref{#1})}
\newcommand{\eqns}[2]{eqs.~(\ref{#1})-(\ref{#2})}

 \newcommand{\beq}{\begin{equation}}
 \newcommand{\eeq}{\end{equation}}
 \newcommand{\beqa}{\begin{eqnarray}}
 \newcommand{\eeqa}{\end{eqnarray}}

\def\bma#1{\mbox{\boldmath{$#1$}}}

 \def\lsim{\ \rlap{\raise 2pt \hbox{$<$}}{\lower 2pt \hbox{$\sim$}}\ }
 \def\gsim{\ \rlap{\raise 2pt \hbox{$>$}}{\lower 2pt \hbox{$\sim$}}\ }

\def\bear{\begin{eqnarray}}
\def\eear{\end{eqnarray}}

 \def\chiu{{\raise 2pt \hbox{$\,\chi$}}{\lower 1pt \hbox{$_{u\,}$}}}
 \def\chiuqt{{\raise 2pt \hbox{$\,\chi$}}
            \hbox{$^4\!\!$}{\lower 1pt \hbox{$_{\!u\,}$}}}
\def\chiusq{{\raise 2pt \hbox{$\,\chi^2\!\!$}}{\lower 1pt \hbox{$_{u\,}$}}} 
\def\chid{{\raise 2pt \hbox{$\,\chi$}}{\lower 1pt \hbox{$_{d\,}$}}}
 \def\chidsq{{\raise 2pt \hbox{$\,\chi^2\!\!$}}{\lower 1pt \hbox{$_{d\,}$}}}
 \def\chie{{\raise 2pt \hbox{$\,\chi$}}{\lower 1pt \hbox{$_{e\,}$}}}
 \def\chiesq{{\raise 2pt \hbox{$\,\chi^2\!\!$}}{\lower 1pt \hbox{$_{e\,}$}}}
 \def\chiud{{\raise 2pt \hbox{$\,\chi$}}{\lower 1pt \hbox{$_{u,d\,}$}}}
 \def\chiude{{\raise 2pt \hbox{$\,\chi$}}{\lower 1pt \hbox{$_{u,d,e\,}$}}}

   \vspace{-2.0cm} 
 \title{
   Yukawa hierarchies from spontaneous breaking of the
    $\bma{SU(3)_L\times SU(3)_R}$ flavour symmetry?}

\author[a,b]{Jos\'e R. Espinosa,}
\author[c]{Chee Sheng Fong}
\author[c]{and Enrico Nardi}

\emailAdd{jose.espinosa@cern.ch}
\emailAdd{chee.sheng.fong@lnf.infn.it} 
\emailAdd{enrico.nardi@lnf.infn.it} 

\affiliation[a]{IFAE, Universitat Aut\`onoma de Barcelona, 08193 
Bellaterra, Barcelona, Spain}
\affiliation[b]{
ICREA, Instituci\`o Catalana de Recerca i Estudis 
Avan\c cats, Barcelona, Spain}
\affiliation[c]{
INFN, Laboratori Nazionali di Frascati,
Via Enrico Fermi 40,    I-00044 Frascati, Italy}

\abstract{
  The tree level potential for a  scalar   multiplet of `Yukawa fields' $Y$     
  for one type of quarks   admits the promising vacuum configuration 
  $\langle Y\rangle \propto {\rm diag}(0,0,1)$  that breaks spontaneously    
  $SU(3)_L\times SU(3)_R$ flavour  symmetry.
  We investigate whether the vanishing entries could be lifted to
  nonvanishing values by slightly perturbing the potential, thus
  providing a mechanism to generate the Yukawa hierarchies.  For
  theories where at the lowest order the only massless states are
  Nambu-Goldstone bosons we find, as a general result, that the
  structure of the tree-level vacuum is perturbatively stable against
  corrections from scalar loops or higher dimensional operators.  We
  discuss the reasons for this stability, and give an explicit
  illustration in the case of loop corrections by direct computation
  of the one-loop effective potential of Yukawa fields.  Nevertheless,
  a hierarchical configuration $\langle Y\rangle\propto {\rm
    diag}(\epsilon',\epsilon,1)$ (with $\epsilon',\, \epsilon\ll 1$)
  can be generated by enlarging the scalar Yukawa sector. We present a
  simple model in which spontaneous breaking of the flavour symmetry
  can give rise to the fermion mass hierarchies.  }


\begin{document}
   \maketitle
\flushbottom


\section{Introduction}

Fermion family replication represents probably the oldest unexplained
puzzle in elementary particle physics, dating back to the discovery of
the muon by Anderson and Neddermeyer at Caltech in 1936.  With the
discovery of all the other second and third generation particles, the
puzzle became even more intriguing because fermions with the same
$SU(3)_C \times SU(2)_L\times U(1)_Y$ quantum numbers have been found
with mass values that span up to five orders of magnitude.
Explaining such strongly hierarchical mass patterns requires a
more fundamental theory than the Standard Model (SM), and  a plethora of
attempts in this direction have been tried. In their
large majority they basically follow two types of approaches:

\begin{enumerate}
\item[(i)] The first is to postulate  new symmetries under
  which fermions with the same SM quantum numbers transform
  differently. The fact that fermion families appear to
  replicate is then just an illusory feature of the low energy theory,
  due to our incomplete knowledge of the full set of
  fundamental quantum numbers.  This is, for example, the basic
  ingredient of the popular Froggatt-Nielsen
  mechanism~\cite{Froggatt:1978nt}, in which the hierarchy of the Yukawa
  couplings follows from a dimensional hierarchy in the corresponding
  effective Yukawa operators,  obtained by assigning to the
  lighter generations larger values of new Abelian charges.

\item[(ii)] A different approach is to assume that the different
  generations contain exact replica of the same set of states. The
  gauge invariant kinetic terms of each type of fermions of the same
  charge and chirality is then characterized at the fundamental level
  by a $U(3)$ (flavour) symmetry~\cite{MFV}.  This symmetry must be
  broken: when the breaking is explicit and provided simply by the
  Yukawa terms we have the SM. However, interesting theoretical
  attempts have been put forth in which the symmetry is broken
  spontaneously by vacuum expectation values (vevs) of scalar `Yukawa'
  fields, that transform under the various $U(3)$ in such a way that,
  at the Lagrangian level, the flavour symmetry is
  exact~\cite{Anselm:1996jm,Berezhiani:2001mh,Koide:2008qm,Koide:2008tr,%
Koide:2012fw,Feldmann:2009dc,Albrecht:2010xh,Grinstein:2010ve,%
Alonso:2011yg,Nardi:2011st,Mohapatra:2012km}.
\end{enumerate}

The first approach basically relies on {\it ad hoc} assignments of new
quantum numbers in order to reproduce qualitatively the observed mass
patterns.  The other approach, which is the one pursued in
this paper, can be considered theoretically more ambitious (as 
it relies  on less {\it ad hoc} assumptions) although it is by
far more challenging than the first one regarding 
successful model implementations. In order to offer a natural solution
to the Yukawa hierarchy, such models should not rely on a hierarchical
arrangement of parameters or some tuning between them while, {\it e.g.},
loop-induced hierarchies would be plausible. 

\section{Symmetry, invariants, and the tree-level scalar potential}

The SM fermions are arranged into triplets of states with the same
gauge quantum numbers, and it is then natural to postulate some
symmetry group that commutes with the SM gauge group and has
three-dimensional representations.  The symmetry, however, is not
realized in the spectrum, and generally this signals a non invariant
ground state yielding spontaneous symmetry breaking (SSB).

The interesting question is which type of SSB, if any, could split the
masses of the members of a multiplet and produce the large hierarchies
that we observe. A step in this direction was taken in
ref.~\cite{Nardi:2011st} and, in order to introduce the theoretical
framework and notation, we will now recall the main results obtained,
following ref.~\cite{Nardi:2011st} in particular.

We do not attempt to build a complete flavour model but rather explore 
the possibilities of this kind of approaches.
We therefore start simply with a pair of flavour triplets of SM 
fermions with opposite chirality $\psi_L^i$, $\psi^j_R$,
($i,\,j=1,2,3$) and in representations of the gauge group such that
their bilinear combination $\bar \psi_L^i \psi^j_R$ can be coupled to
the Higgs in a gauge invariant way.
The largest symmetry of their  gauge invariant kinetic term
is $U(3)_L\times U(3)_R$ where the first factor acts on the
electroweak fermion doublets $\psi_L$ and the second on the weak
singlets $\psi_R$.  Here we will concentrate on the semisimple flavour
subgroup
\beq
\label{eq:GF}
{\cal G}_F = SU(3)_L\times SU(3)_R
\eeq
since the fate of the $U(1)$ factors (whether they are broken or
contribute to linear combinations of unbroken generators, as
e.g. Baryon number) is of no relevance in what follows.
We assume that the SM Yukawa term which couples $\psi_{L,R}$ to the Higgs 
field $H$ originates from a non-renormalizable effective coupling
\begin{equation}
  \label{eq:nonren}
  {\cal L}_Y = \frac{1}{\Lambda}\, \bar \psi_L\, Y\, \psi_R\, H 
\end{equation}
that involves a scalar `Yukawa' field $Y$ (which is, in fact, a matrix
in flavour space)   and    some large scale $\Lambda$  at which the effective
Yukawa operator arises.  Invariance of ${\cal L}_Y$ under ${\cal G}_{F}$
fixes the following quantum number assignments under $SU(3)_L\times SU(3)_R$:
\begin{eqnarray}
  \label{eq:assinments}
  \psi_L \sim (3,1), \qquad 
  \psi_R \sim (1,3), \qquad 
  Y \sim (3,\bar 3)\,. 
\end{eqnarray}
If $Y$ acquires a vev, the flavour group ${\cal G}_F$ gets
spontaneously broken. This of course amounts to interpreting the SM
explicit breaking as the result of SSB.

\mathversion{bold}
\subsection{The $T, A,  {\cal D}$ invariants}
\mathversion{normal}

In the following we will denote by $Y$ a generic background field
configuration with components of constant value, although sometimes the
spacetime dependence will be indicated explicitly, $Y(x)$, to emphasize 
this is a field (matrix). Configurations that minimize the potential will be 
instead denoted by $\langle Y \rangle$.
To write down the most general renormalizable ${\cal G}_F$-invariant
potential for $Y$, and explore its possible ground state
configurations and properties, let us consider the
characteristic equation for the eigenvalues $\xi$ 
of the Hermitian matrix $YY^\dagger$
\begin{eqnarray}
  \label{eq:secular}
{\cal P}(\xi)\equiv \det\left(\xi I- Y Y^{\dagger}\right)= \xi^3
- T\, \xi^2 + A\, \xi  - D^2=0\,, 
\end{eqnarray}
where $I=I_{3 \times 3}$ is the identity matrix in flavour space, and
the coefficients are
\begin{eqnarray}
\label{eq:T}
  T &=& {\rm Tr}( YY^\dagger)= \sum_i \xi_i\,, \\
\label{eq:A}
  A &=& {\rm Tr}\left[{\rm Adj}( YY^\dagger)\right]= 
\sum_{i> j} \xi_i\xi_j\,,\\  
\label{eq:D2}
  D^2 &=& {\rm Det}(YY^\dagger)= \prod_i \xi_i \,. 
\end{eqnarray}
Being the eigenvalues invariant under group transformations, so are
the coefficients of the characteristic equation for $Y Y^{\dagger}$,
namely its trace $T$ (positive definite and of dimension 2), the trace 
of its adjugate (or equivalently of the cofactor) matrix $A$ (positive
definite and of dimension 4), and its determinant $D^2$, which is an
invariant of dimension 6.  However, under special unitary
transformations $V_{L,R}$ of $SU(3)_{L,R}$ (with ${\rm Det}
V_{L,R}=+1$) we have for the determinant of $Y$: ${\cal D}\equiv {\rm
  Det}(Y)\to {\rm Det}(V_LY V_R^\dagger)= {\cal D}$, so that $\cal D$
is also an invariant, but of dimension 3 and thus renormalizable. We
conclude that $T(x)$, $A(x)$ and ${\cal D}(x)$ are the renormalizable
symmetry invariant field combinations from which 
the scalar potential can be constructed.

In fact, one can show that the most general $SU(3)_L\times SU(3)_R$
invariant potential including nonrenormalizable terms of any
dimension, can always be expressed as a function of just the three
$T,A,{\cal D}$ invariants, that is, it has the form 
$V(T,A,{\cal D},{\cal D}^*)+{\rm h.c.}\,$.
This amounts to proving that any
invariant term of higher order in $Y$ can be reduced to powers of
$T,A,{\cal D}$.  For determinants of higher powers of $Y$ we have
straightforwardly ${\rm Det} ({Y}^m Y^\dagger{}^n)={\cal D}^m {\cal
  D}^*{}^n$. For trace invariants let us define:
\begin{equation}
  \label{eq:TmAm}
 T_{2n}= {\rm Tr}[(YY^\dagger)^n]\,, \qquad\quad 
 A_{2n}= {\rm Tr}[{\rm Adj}(YY^\dagger)^n]\,. 
\end{equation}
According to this notation, $T=T_2$ and $A=A_2$.
It is  straightforward to show that 
\begin{equation}
  \label{eq:T4A4}
 T_{4}= T_2^2-2\, A_2\,,  \qquad\quad A_4=A_2^2-2\,T_2\, D^2\,.
\end{equation}
To show that higher order invariants $T_{2n}\,,A_{2n}$ with $n>2$ can
also be written in terms of $T_2,\,A_2,\,D^2$ we can make use of the
Cayley-Hamilton theorem, which states that every square matrix of
complex numbers satisfies its own characteristic equation. That is, by
substituting $\xi \to YY^\dagger$ in ${\cal P}(\xi)$ of~\eqn{eq:secular},
one has the matrix equation:
\begin{eqnarray}
  \label{eq:secularA}
{\cal P}(YY^\dagger) = 
 (YY^\dagger)^3
- T_2\, (YY^\dagger)^2 + A_2\, (YY^\dagger)  - D^2\, I=0\,. 
\end{eqnarray}
This allows to rewrite $(YY^\dagger)^3$, and thus recursively any
other higher power of $YY^\dagger$, in terms of the three fundamental
invariants and of $(YY^\dagger)$ and $(YY^\dagger)^2$ which in turn
reduce to $T_2,\,A_2,\,D^2$ after taking the trace or after tracing
their adjugates and using~\eqn{eq:T4A4}.

\subsection{Scalar potential and tree-level vacua}

A necessary
condition to ensure that the observed hierarchy of the SM Yukawa
couplings is reproduced, is that, at the SSB minimum:
\begin{equation}
   \label{eq:qualitative}
 \langle D \rangle ^{{1}/{3}} \ll 
\langle A\rangle^{{1}/{4}}  \ll \langle T\rangle^{{1}/{2}}\,,  
\end{equation}
where $D=\left|\cal D\right|$. The first goal is then to construct a
scalar potential which naturally has such minimum.  In terms of the
$T,A,{\cal D}$ invariants the most general renormalizable potential
for $Y$ can be written as~\cite{Nardi:2011st}\footnote{ As long as
  $\langle H^\dagger H \rangle/\Lambda^2 \ll 1$ the coupling with the
  Higgs, $ H^\dagger H T$, can be omitted from \eqn{eq:V3}. Regarding
  the effects of such coupling on the Higgs potential, electroweak
  symmetry breaking at the correct scale would require a certain
  degree of fine-tuning in the term $H^\dagger H\left(\langle
    T\rangle-\mu_H^2\right)$.}
\begin{equation}
  \label{eq:V3}
  V_0 =  V_{T} + V_{A}+V_{\cal D} \,,     
\end{equation}
with
\begin{eqnarray}
  \label{eq:VT3}
V_{T} &=&\lambda \left[T- \frac{m^2}{2\lambda}\right]^2\!, \\
  \label{eq:VA3}
  V_A &=&  \lambda_A A\,, \\ 
  \label{eq:VD3}
V_{\cal D} &=& \tilde \mu\, {\cal D} +  \tilde \mu^*\, {\cal D}^* 
=2\,\mu\,D\, \cos\phi_{\cal D}\,. 
\end{eqnarray}
We assume that all the Lagrangian parameters are evaluated at the
scale $\Lambda$, which can be identified with that in \eqn{eq:nonren}.  
$V_T$ in~\eqn{eq:VT3} contains the two renormalizable invariants
constructed from the trace, $V_T=\lambda T^2 -m^2 T$, plus an irrelevant
constant.  We require $\lambda>0$ and $m^2 > 0$ in order to have a
potential bounded from below and to trigger SSB. The parameter
$\lambda_A$ which multiplies $A$ can be either positive or
negative, and we need to consider both possibilities.  
 The last equality
in~\eqn{eq:VD3} is obtained after defining $\tilde \mu = \mu
e^{i\delta}$ with $\mu\equiv |\tilde \mu|$, ${\cal D}= e^{i\varphi(x)}D$, and $\phi_{\cal D}(x)=\varphi(x)+\delta$.

Let us now seek the most general form for the vev of the scalar field.
A generic $3\times 3$ matrix of (complex) constant background fields $Y$ has 9
moduli and 9 phases, and by means of an $SU(3)_L\times SU(3)_R$
rotation (corresponding to 3+3 moduli and 5+5 phases) can always be
brought into diagonal form $Y^{(d)}={\rm diag}(Y_{11},Y_{22},Y_{33})$.
Since the $SU(3)_{L,R}$ diagonal generators $\lambda_3$ and
$\lambda_8$ commute with $Y^{(d)}$, this matrix is invariant under the
subgroup $U(1)_{(\lambda_3)_{L+R}}\times U(1)_{(\lambda_8)_{L+R}}$ and
therefore, out of the initial 9 phases, only $10-2=8$ can be removed by
flavour rotations. Without loss of generality we can then choose a
basis in which the background classical field has the form
 \begin{equation}
   \label{eq:basis}
Y =\frac{1}{\sqrt{2}}\,{\rm diag}(R_{11},R_{22},R_{33}+i J_{33})\,,
 \end{equation}
where $R_{ii}$ and $J_{ii}$ are real scalar fields. 

Regarding the value of $Y$ that minimizes the
 potential in \eqn{eq:V3}, that is, the tree-level vev of $Y$, from
 \eqn{eq:VD3} we immediately see that if $\langle D \rangle \neq 0$
 then $V_{\cal D}$ is minimized when $\cos\phi_{\cal D}=-1$
(by $\langle \varphi \rangle=\pi-\delta$), 
so that we can restrict our analysis to $D\geq 0$ in what follows. 
If instead $\langle D \rangle = 0$ at the minimum of the potential,
the phase of $\langle {\cal D} \rangle$ is undetermined (and $\varphi$ has a flat potential).  
This allows us to set $\langle\varphi\rangle =\pi-\delta$  
and search for the minimum around the configuration
 \begin{eqnarray}
   \label{eq:VDY1}
   V_{\cal D}^{\rm min} &\equiv& V_D = - 2 \mu D, \\    
   \label{eq:VDY2}
   \langle Y\rangle  &=&\frac{1}{\sqrt{2}}\,{\rm diag}(R_{11},R_{22},R_{33})\,,
 \end{eqnarray}
where, with a slight abuse of notation, we have denoted   with 
$R_{33}$ the modulus $\sqrt{2}|Y_{33}|$.
 From \eqn{eq:VT3} we immediately see that $V_T$ is minimized on the
 surface $\langle T\rangle= m^2/(2\lambda)$ of the sphere in the
 eighteen dimensional parameter space.  Note that since we must
 require $m^2/(2\lambda)\lsim \Lambda^2$ in order to explain e.g. the
 value of the top-quark Yukawa coupling\footnote{For example, at a cutoff
   scale $\Lambda \sim (10^9 - 10^{12})\,$GeV we have
   $y_t=\langle Y_{33}\rangle/\Lambda \sim
   0.6-0.5$~\cite{Xing:2007fb}. In particular, this justifies
   neglecting contributions to the Yukawa terms~\eqn{eq:nonren} of
   dimension higher than five.}, then a perturbative
 $\lambda<1$ implies $m^2 <\Lambda^2$, consistently with the effective
theory treatment. Concerning $A$ and $D$, they are both
 maximized for symmetric vacua $\langle Y \rangle \propto   
 \, {\rm diag}(1,\,1,\,1) $ 
and their minimum value is zero.  
To ensure $\langle D\rangle=0$, at
 least one entry in $\langle Y \rangle$ must vanish, while for
 $\langle A\rangle =0$ two entries must vanish, e.g.  $\langle Y
 \rangle \propto \, {\rm diag}(0,\,0,\,1) $. 
 Which particular minimum on
 the surface of constant $\langle T \rangle$ is selected depends on the
 sign and value of $\lambda_A$ and on the value of $\mu$.
 Following~\cite{Nardi:2011st} we recall below which types of SSB
 minima can occur and under which conditions.

\begin{itemize}

\item[(i)] When $\lambda_A<0$  we have to require
  $|\lambda_A|<3\lambda$ in order that the potential remains bounded from
  below.  $A$ is maximized for symmetric vacua (\eqn{eq:vYs} below)
  and since $V_A=\lambda_A A$ is negative this is the favoured
  configuration.  $D$ is also maximized for symmetric vacua so that
  the negative value of $V_{\cal D}$,~\eqn{eq:VDY1}, further lowers the
  minimum. The symmetric vev
  \begin{equation}
    \label{eq:vYs}
    \langle Y\rangle^{s} = 
v_s\, {\rm diag}(1,\,1,\,1),    
  \end{equation}
  (where $v_s$ is given in eq.~(18) of~\cite{Nardi:2011st} with
  $\lambda'=-\lambda_A$) 
%
%
%
%
  corresponds, however, to non-hierarchical  Yukawa couplings
  yielding $\langle T\rangle \approx 
\langle A\rangle^{1/2} \approx \langle
  D\rangle^{2/3} = v_s^2$.

\smallskip

\item[(ii)] When $\lambda_A>0$, $V_A=\lambda_A A$ is always positive
  and  minimized for $\langle A \rangle=0$, which favours
  (hierarchical) vacua with two vanishing entries
  \begin{equation}
    \label{eq:vYh}
    \langle Y\rangle^{h}  =v_h\, {\rm diag}(0,\,0,\,1)  \,.  
  \end{equation}
with  $v_h=m/\sqrt{2\lambda}$.
  Given that $V\left(\langle Y\rangle^{h}\right)=0$, this configuration is
  selected as long as the potential in the symmetric direction has a
  positive definite value $V\left(\langle Y\rangle^{s}\right)>0$ 
  in spite of a possible negative contribution from $V_D$. This occurs
  as long as (see~\cite{Nardi:2011st} for details)
\begin{equation}
  \label{eq:sol3}
  \frac{\mu^2}{m^2}<2 \lambda
\left[\left(4+\frac{\lambda_A}{\lambda}\right)^{3/2}
-\left(8+3\frac{\lambda_A}{\lambda}\right)\right]\,. 
\end{equation}
In this case $\langle T\rangle = v_h^2$ while $\langle D\rangle
=\langle A\rangle =0$, which represents a promising first approximation
to the realistic hierarchy, as in~\eqn{eq:qualitative}.

\end{itemize} 

Thus,  the tree level analysis indicates that the most general
renormalizable ${\cal G}_F=SU(3)_L\times SU(3)_R$ invariant potential
admits two types of SSB vevs, that lead to the two symmetry breaking
patterns ${\cal G}_F\to H_s$ and ${\cal G}_F\to H_h$ respectively 
with little groups:  
\begin{eqnarray}
  \label{eq:Hs}
  H_s &=& SU(3)_{L+R}\,, \\
  \label{eq:Hh}
  H_h &=& SU(2)_L\times SU(2)_R\times U(1)_{(\lambda_8)_{L+R}}\,.  
\end{eqnarray}

Before concluding this section let us recall some jargon specific to
SSB problems, as well as some general results. The largest subgroup 
$H\subset {\cal G}_F$ that leaves invariant some background field
configuration $\langle Y\rangle$ is called the {\it little group} of
$\langle Y\rangle$.  In particular, the little groups $H_{s,h}$ in
\eqns{eq:Hs}{eq:Hh} are {\it maximal little groups} in the sense that
none is contained in the other or in another little group of ${\cal
  G}_F$.  Acting with group elements in ${\cal G}_F/H$ on $\langle
Y\rangle$ while keeping its `length' ($\langle T\rangle$ in our case) 
fixed, one obtains the {\it orbit} of $\langle Y\rangle$.  If in the
neighbourhood of $\langle Y\rangle$ all other background
configurations have the same little group $H$, the collection of their
orbits is called an {\it open stratum} (or {\it dense stratum}). An
example of a configuration with orbit belonging to an open stratum is
the general form $\langle Y \rangle \sim {\rm diag}(a,b,c)$ 
corresponding to~\eqn{eq:VDY2}, with little group\footnote{Considering the 
full initial group of invariance
  $U(3)^2=U(1)_B\times U(1)_{L-R}\times SU(3)^2$ one readily
  recognizes that the full little group of continuous transformations
  for the Yukawa configuration~\eqn{eq:VDY2} is $=U(1)_B\times
  U(1)_{(\lambda_3)_{L+R}}\times U(1)_{(\lambda_8)_{L+R}}$ which are
  linear combinations of the three baryon flavour symmetries
  $U(1)_{B_i}$ (eventually broken to $U(1)_B$ by
  inter-generational quark mixing).
}
\beq
  \label{eq:Hg}
  H_{abc}=U(1)_{(\lambda_3)_{L+R}}\times U(1)_{(\lambda_8)_{L+R}} \,.  
\eeq
%
The boundaries of an open stratum are the closed strata, which contain 
orbits in the neighborhood of which there are other configurations with
different little groups.
%
If $H \subset {\cal G}_F$ is a maximal little group, then the
corresponding $\langle Y\rangle$ is in a closed stratum.  If the
closed stratum has only one orbit, this orbit is a stationary point of
any smooth real invariant functions of $\langle Y\rangle$
(Michel-Radicati theorem~\cite{Michel:1971th}).

In our case, the boundaries of the open stratum of the general
background field configuration in~\eqn{eq:VDY2} are $\langle
Y\rangle^{s,h}$ in~\eqns{eq:vYs}{eq:vYh} and, in agreement with
Michel's theorem~\cite{Michel-conj}, their little groups
$H_{s,h}$~[\eqns{eq:Hs}{eq:Hh}] are the {\it maximal stability groups}
of the most general fourth order function of the invariants (the
tree-level potential).

\begin{figure}[t]
\centering
\includegraphics[scale=1.2]{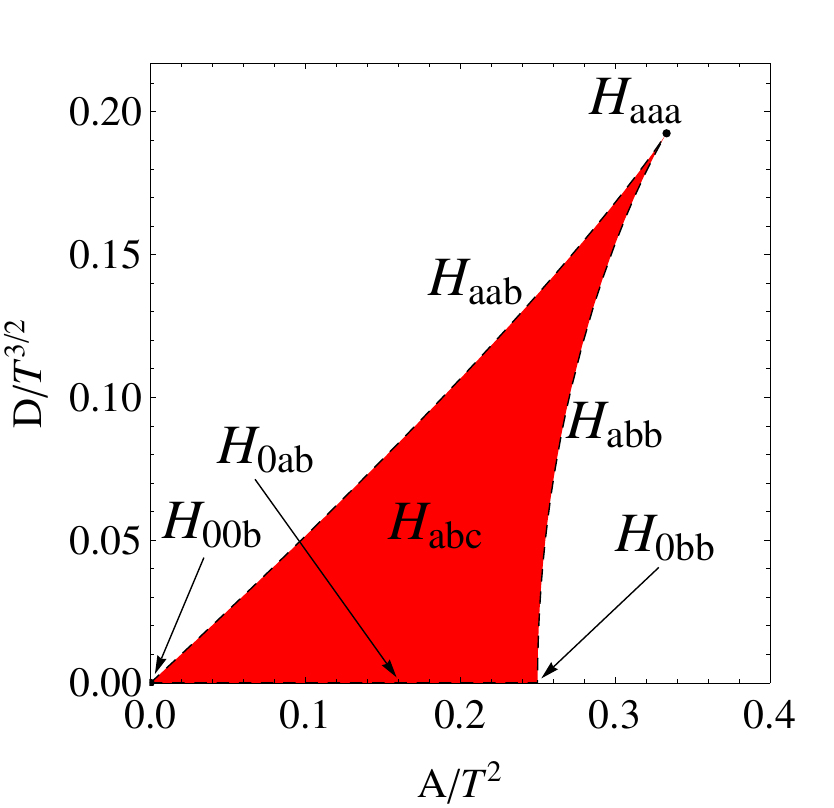}
\caption{The orbit space for the group $SU(3)_L\times SU(3)_R$ broken
  by a field $Y$ in the bi-fundamental representation, plotted in the
  plane $(A/T^2, D/T^{3/2})$ with an arbitrary nonzero $T$.  The $H$'s
  labeling the points in the plot denote the little groups of the
  corresponding field configurations $\langle Y\rangle$.
  $H_{aaa}=H_s$ and $H_{00b}=H_h$ are the maximal little groups of $
  \langle Y\rangle\sim{\rm diag}(a,a,a)$ and $ \langle
  Y\rangle\sim{\rm diag}(0,0,b)$.  }
\label{orbitspace} 
\end{figure}

As we have seen, although the potential is a function of many scalar
fields ($Y$ has 18 degrees of freedom) the fact that only the 3
invariants $T$, $A$ and $D$ enter, simplifies greatly the minimization
problem. A further simplification follows from the observation
\cite{JSKim} that the orbit space can be described in a compact way by
$T$ (which is nonzero in any symmetry breaking vacuum) and by the two
dimensionless ratios of invariants $r_A\equiv A/T^2$ and $r_D\equiv
D/T^{3/2}$, which are respectively bounded within the intervals $0\leq
r_A\leq 1/3$ and $0\leq r_D\leq 1/(3\sqrt{3})$.
For  given  fixed values of $r_A$ and $r_D$, 
the potential 
is a function of $T$ only, with some minima $\langle T\rangle
(r_A,r_D)$. The global minimum is then the deepest of such minima in
the whole range ($r_A,r_D$).  Such an orbit space does not depend on
the details of the potential, but only on the group structure and on the
representation of the scalar fields.  
For our case the orbit space is plotted in
figure~\ref{orbitspace} in which, for instance, $H_{aaa}$ refers to
the symmetric little group $H_s=SU(3)_{L+R}$ and labels the field
configuration $\langle Y\rangle\sim{\rm diag}(a,a,a)$, while a generic
point $H_{abc}$ labels the field configuration $\langle
Y\rangle\sim{\rm diag}(a,b,c)$.  For our labels we keep the ordering
$a<b<c$, so that the plot gives also information on the relative size
of the $\langle Y\rangle$ entries.
Note that the point labeled $H_{0bb}$ belongs to the same stratum as
$H_{abb}$, while the point $H_{0ab}$ belongs to the same stratum as
$H_{abc}$.
Regarding $H_{aab}$ and $H_{abb}$, although they label disconnected
regions in orbit space,  they correspond to the same little group,
that is $SU(2)_{L+R}\times  U(1)_{(\lambda_8)_{L+R}}$,
and so they belong to the same stratum.
Following the method proposed in ref.~\cite{JSKim}, the little groups
of the global minima of $V_0$ can also be determined through an
analysis of equipotential lines in the orbit space.  The result is
that for $\lambda_A < 0$ the minimum of $V_0$ always is at
$H_{aaa}=\left(1/3,1/(3\sqrt{3})\right)$, while for
$\lambda_A > 0$ the minimum can be at $H_{00b}=(0,0)$ or at
$H_{aaa}$, depending on the values of the parameters (i.e.
if~\eqn{eq:sol3} is satisfied or not).  This of course coincides with
our previous findings.

The question now is if, as suggested by Michel's
conjecture~\cite{Michel-conj}, the one-loop effective potential can
only have minima with the same maximal little groups $H_{s,h}$ as the
tree-level potential, or if phenomenologically more appealing minima
with the generic little group $H_{abc}$ are possible.  On one
hand, if a phenomenologically acceptable minimum had corresponded to a
maximal little group, it would have been easy to model a potential
leading to it, as this is facilitated by the group structure which
makes such symmetry breaking natural. On the other hand, seeking for a
symmetry breaking pattern with non-maximal little group is harder but
might offer some deeper insight into the dynamics producing such
breaking. Be that as it may, as we will discuss in the next two
sections, although counter-examples to Michel's conjecture are known
to exist~\cite{Michel-counterex}, in our case $H_{s,h}$ remain stable
with respect to perturbative effects from loop corrections or from
operators of higher dimension.

\section{Sequential breaking}
\label{sec:loop}

The symmetric solution $\langle Y\rangle^{s}= v_s\,
\mathrm{diag}(1,1,1)$ in~\eqn{eq:vYs}, which yields non-hierarchical
Yukawa couplings, is phenomenologically uninteresting and we do not
consider it (although it might be of interest in the neutrino sector).
The solution $\langle Y\rangle^{h} = v_h\, \mathrm{diag}(0,0,1)$~in
\eqn{eq:vYh} appears instead as a promising first approximation to the
observed Yukawa hierarchies, but is tenable only if the two vanishing
entries can be lifted to small nonzero values by some effect. In
this section we want to consider the possibility of obtaining a
vacuum $\langle Y\rangle=  v_Y\, {\rm
  diag}(\epsilon',\epsilon,1)$ with the non-maximal little group
$H_{abc}\subset H_h$, from a small perturbation of the tree-level vacuum
$\langle Y\rangle^{h}$ in~\eqn{eq:vYh}.
Such perturbation could be provided, for instance, by one-loop
corrections to the effective potential $V_{\rm eff}$ (as hypothesized
in ref.~\cite{Nardi:2011st}) or by higher dimensional operators. We
will show that in general these possibilities cannot be realized.

Let us focus for the moment on the symmetry breaking pattern ${\cal
  G}_F \to H_h$, {\it i.e.} the minimum is $\langle Y\rangle^h =v_h\,
{\mathrm{diag}}(0,0,1)$ with $v_h=m/\sqrt{2\lambda}$.  There are
$n_G=8+8$ generators in ${\cal G}_F$, and $n_H=3+3+1=7$ unbroken
generators in the little group $H_h$. Nine generators are thus broken
and accordingly we find in the spectrum $n_b=9$ massless
Nambu-Goldstone Bosons (NGB).  The remaining nine states are arranged
in multiplets of $H_h$ with masses:
\begin{eqnarray}
  \label{eq:eigen1}
   m^2_1 &=& 
4\lambda v_h^2 \,, 
\hspace{2.4cm} \qquad  {\rm 1\ state}, \\
  \label{eq:eigen+-}
  m^2_\pm &=& 
\lambda_A v_h^2 \pm\mu v_h \,, 
\hspace{0.4cm}\qquad  4+4{\rm\ states}.
\end{eqnarray}
Assuming that upon minimization of the 
one-loop effective potential a ground state with little group $H'\subset
H_h$ with $n_{H'} < n_{H_h}$ unbroken and $n_{b'}> n_b$ broken generators is
obtained, would then imply that out of the 9 massive states,
$n_{H_h} - n_{H'}$ have to become the new massless NGB. The
required cancellation between the tree-level and loop mass
contributions do not conform to the perturbative approach to this problem.  
In other words, the
zeroth-order mass relations determined by the unbroken subgroup of the
symmetry group should persist also at higher
orders~\cite{Georgi:1972mc}. More precisely, a theorem proved long ago
by Georgi and Pais~\cite{Georgi:1977hm} states that a reduction of the
tree-level vacuum symmetry via radiative corrections can only occur if
there are additional massless bosons in the tree
approximation.\footnote{Perhaps a bit surprisingly, this applies also
  to radiative breaking of discrete symmetries like
  $CP$~\cite{Georgi:1977hm}.}  
This can be seen in the following
 way: the Goldstone theorem states that for each generator ${\cal
   T}^a$ of a continuous symmetry acting nontrivially on the vacuum
 there is a massless
 scalar, that is:
\begin{equation}
{\cal T}^a \cdot \langle Y\rangle\neq 0  \implies
M^2\cdot {\cal T}^a\cdot \langle Y\rangle =0\,, 
\label{eq:goldstone} 
\end{equation}
where $M^2$ is the second derivative of $V$ with respect to the
fields evaluated at $Y= \langle Y\rangle $.  
Following ref.~\cite{Georgi:1977hm},   
let us now write $M^2
= M^2_0 + \delta M^2$ and $ \langle Y\rangle = \langle Y\rangle_0 +
\delta \langle Y\rangle$ where $M^2_0$ and $ \langle Y\rangle_0$ are
obtained from minimization of $V_0\,$, while $\delta M^2$ and $\delta \langle
Y\rangle$ are the perturbations induced by higher order corrections
to the potential.  
Stepwise breaking implies that there must be some 
generator ${\cal T}^a$ in the little group $H$ of the tree-level 
vacuum that is not in $H'$, that is:
\begin{equation}
  \label{eq:GP}
{\cal T}^a \cdot \langle Y\rangle_0 = 0, \qquad {\rm and} 
\qquad   {\cal T}^a \cdot \langle Y\rangle \neq 0\,. 
\end{equation}
At first order in the corrections,~\eqn{eq:goldstone} becomes
\begin{equation}
M^2_0\cdot {\cal T}^a \cdot \langle Y\rangle_0 + 
\delta M^2\cdot {\cal T}^a \cdot \langle Y\rangle_0 + 
M^2_0 \cdot {\cal T}^a\cdot \delta \langle Y\rangle 
=0\,. 
\label{eq:goldstone2} 
\end{equation}
The first two terms in this equation vanish because of the first
equation in~(\ref{eq:GP}), and the vanishing of the last term then
implies additional massless scalars, that at the tree level are not
NGB. In this case, a further breaking of the
symmetry by higher order effects can simply transform some of these
additional massless states into NGB.  A well known example is the
Coleman-Weinberg (CW) potential~\cite{Coleman:1973jx} in which the
scale invariance at the tree level implies (non-Goldstone) massless
states in the lowest order approximation.  The breaking of the
symmetry at one loop then transforms some massless states
into NGB and gives mass to the remaining ones.

A crucial point in our particular case is that the extra massless
scalars in the tree approximation should appear {\it naturally}, {\it
  e.g.}  due to some extended accidental symmetry, because of some
larger symmetry of the tree-level vacuum~\cite{Georgi:1975tz} or due
to renormalization group evolution of parameters (so that these
massless scalars appear at some particular renormalization scale).
For example, it is technically possible to tune the mass $m^2_-$ in
\eqn{eq:eigen+-} to zero by setting $\lambda_A m=\mu\sqrt{2\lambda}$,
and this would result in four additional massless states at the tree
level. To verify if this condition allows for a further breaking of
the symmetry by loop corrections, we have carried out a numerical
minimization of the $SU(3)_L\times SU(3)_R$ one-loop effective
potential given in the Appendix.  In the general case with no
additional massless states, the parameter space of the effective
potential remains divided into two regions corresponding to the two
vacuum structures $\langle Y\rangle^h = v_h\, \mathrm{diag} (0,0,1)$
and $\langle Y\rangle^s = v_s\, \mathrm{diag}(1,1,1)$ with a boundary
that is still given with a good approximation by
condition~\eqn{eq:sol3}.  By setting $\lambda_A m=\mu\sqrt{2\lambda}$
one obtains that this condition is never satisfied,  the vacuum
remains in the symmetric configuration $\langle Y\rangle^s$ with
little group $H_s$ and no further breaking occurs.

Although there is no natural way to forbid the two terms $V_A$ and
$V_{\cal D}$ in the potential, additional massless states would be
obtained by setting by hand $\mu\to 0$ and $\lambda_A\to 0$ in the
tree-level potential~\eqn{eq:V3}.  This is because the  surviving part $V_T$,
\eqn{eq:VT3}, has an accidental symmetry $SO(18)$ that is much larger
than $SU(3)_L\times SU(3)_R$. The minimum condition $T=\sum_i
(YY^\dagger)_{ii}={m^2}/{(2\lambda)}$ fixes the radius of the
eighteen-dimensional hypersphere, leaving the vacuum symmetry $SO(17)$
of the hypersurface which is broken by a choice of the vacuum
direction.  We thus obtain seventeen massless bosons: the nine NGB of
the broken $SU(3)_L\times SU(3)_R$ plus $4+4$ additional massless
states corresponding to the eigenvalues in~\eqn{eq:eigen+-}. Clearly,
in this case since the symmetry of the full theory is just
$SU(3)_L\times SU(3)_R$ of the Yukawa operator~\eqn{eq:nonren}, it is
conceivable that corrections due to other interactions (as the ones
that give rise to the effective Yukawa operator, or $SU(3)_L\times
SU(3)_R$ gauge interactions if the flavour symmetry is gauged) could
eventually yield a stepwise breaking.  We will discuss further this
possibility in the next section.

In conclusion, lifting the tree-level vev $\langle Y \rangle^h= v_h\,  
{\rm diag}(0,\,0,\, 1)$ to a vev with hierarchical components $\langle
Y \rangle =v_Y\,  {\rm diag}(\epsilon',\, \epsilon,\, 1)$ would
require at least the symmetry reduction $SU(2)_L\times SU(2)_R\times
U(1)_{(\lambda_8)_{L+R}} \to U(1)_{(\lambda_3)_{L+R}}\times
U(1)_{(\lambda_8)_{L+R}}$. If, as in the case we are considering,
there are no additional tree-level massless states, the $n_{H_h}-n_{H_{abc}}=5$
new NGB cannot appear at the loop level, and thus no stepwise
reduction of the symmetry can occur. We will mention in
Section \ref{sec:generalization} possible loopholes to this
conclusion.

Concerning the possibility that corrections from higher dimensional
operators~\cite{Feldmann:2009dc} could provide the sequential symmetry
breaking we are looking for, we can again rely on the Goldstone
theorem to formulate the necessary conditions for this to happen.
Including higher dimensional terms, we can write the scalar potential
as
\begin{eqnarray}
   \label{eq:dimensional}
V(p_i,\bar\Lambda) &=& V_0(p_i) + 
\sum_k \frac{1}{{\bar\Lambda}^{k}}\,{\cal O}^{4+k}\,,    
 \end{eqnarray}
 where $V_0(p_i)$ is the tree-level potential,~\eqn{eq:V3}, which
 depends on the parameters $\{p_i\}=\{m,\,\mu,\,\lambda,\,\lambda_A\}$
 and ${\cal O}^{4+k}$ represents the set of operators with dimension
 $4+k$ ($k\geq 1$) that are perturbatively suppressed by powers of the
 high energy scale $\bar\Lambda$.
\Eqn{eq:goldstone2} translates straightforwardly into 
\begin{equation}
  \label{eq:barLambda}
 M^2_0(p_i) \cdot {\cal T}^a\cdot \delta \langle Y\rangle 
=0\,. 
\end{equation}
where $\delta \langle Y\rangle$ depends on $\bar\Lambda$.  Stepwise
breaking implies that $ {\cal T}^a\cdot \delta \langle Y\rangle$ is
non vanishing, which in turn requires additional non NGB massless
states in the tree level mass matrix $M^2_0(p_i)$. In the absence of
these states, we can then conclude that also operators of higher
dimension are unable to induce perturbatively sequential symmetry
breaking.

\section{Generalization}
\label{sec:generalization}

We have seen that the maximal little groups $H_{s,h}$ of the stationary
points of the tree-level potential,~\eqn{eq:V3}, 
are stable against corrections to the effective potential induced by
loops or higher dimensional operators. Thus, hierarchical minima with the
observed structure $\langle Y \rangle = v_Y\, {\rm diag}(\epsilon',\,
\epsilon,\, 1)$ cannot simply occur as a perturbation from 
$\langle Y \rangle^h= v_h\, {\rm diag}(0,\,0,\, 1)$. However,
ref.~\cite{Nardi:2011st} found that, if the one-loop potential were to have
terms like $A \log A$ and $D \log D$, they could produce
global minima different from $\langle Y\rangle^h$ or $\langle Y\rangle^s$
in~\eqns{eq:vYs}{eq:vYh}.  Also, a potential
 that indeed can have a hierarchical minimum $\langle Y
\rangle  = v_Y\,{\rm diag}(\epsilon',\, \epsilon,\, 1)$ is given in
appendix B.2 of ref.~\cite{Alonso:2011yg}. Such examples show
that there is no group theory 
obstruction to finding potentials
with minima that have little groups different from the maximal
ones $H_{s,h}$.
  
We will now show that under some general conditions the smallest
little group preserved by the minima of smooth functions of the
$SU(3)_L\times SU(3)_R$ invariants is $SU(2)_{L+R}\times
U(1)_{(\lambda_8)_{L+R}}$. This corresponds, in the diagonal basis, to
the structure $\langle Y\rangle \sim {\rm diag}(a,a,b)$, which can only
allow for a partial hierarchy.  In contrast with the results presented
in Section~\ref{sec:loop}, the present argument
does not rely on perturbative expansions, and can be applied to more general
classes of effects.

Let us consider from now on a generic smooth potential function
$V(T,A,D)$, where we take ${\cal D}$ real from the start and, 
without loss of generality, the invariants are written in the
real diagonal basis for $Y$, as in \eqn{eq:VDY2}.  
 The minimization equations are ($i=\{1,2,3\}$)
\begin{equation}
  \label{eq:minimization}
  \frac{\partial {V}}{\partial T}
\frac{\partial T}{\partial \xi_i}  +
\frac{\partial {V}}{\partial A}
\frac{\partial A}{\partial \xi_i} +
\frac{\partial {V}}{\partial D}
\frac{\partial D}{\partial \xi_i}  = 0\,,
\end{equation}
where, as above, the $\xi_i$ are the eigenvalues of $\langle
YY^\dagger\rangle$ (i.e.  $\langle YY^\dagger\rangle={\rm
  diag}(\xi_1,\xi_2,\xi_3)$ in our specific basis).  We have
\begin{equation}
  \label{eq:derv}
  \frac{\partial T}{\partial \xi_i} =1\,, \qquad
\frac{\partial A}{\partial \xi_i} = T-\xi_i\,, \qquad
\frac{\partial D}{\partial \xi_i} = \frac{1}{2}\,\frac{ D}{\xi_i}\,,
\end{equation}
so that~(\ref{eq:minimization}) leads to the
quadratic equation in $\xi_i$:
\begin{equation}
  \label{eq:minimization2}
\frac{\partial {V}}{\partial A} \, 
\xi_i^2 -
 \left[\frac{\partial {V}}{\partial T}
  + \frac{\partial {V}}{\partial A}\, T \right]\xi_i 
- \frac{D}{2}\,\frac{\partial {V}}{\partial D}
= 0\,.
\end{equation}
Thus, at the extremal point, the three $\xi_i$ eigenvalues should be
solutions of the same equation above which, being quadratic, can only
have at most two different roots. It is in this sense that we expect
generically to have a little group at least as large as
$SU(2)_{L+R}\times U(1)_{(\lambda_8)_{L+R}}$.
From this, it follows that regardless of the type of potential
function ${V}$, the only way to obtain three distinct values
$\xi_1, \xi_2, \xi_3$ is to make zero the three coefficients in
\eqn{eq:minimization2}. In this particular case 
the minimization equations reduce to:  
\beq
  \label{simplemin}
  \frac{\partial {V}}{\partial T}
=0\ ,\;\;\;
\frac{\partial {V}}{\partial A}
=0\ ,\;\;\;
\frac{\partial {V}}{\partial D} =0\ .
\eeq
We could have anticipated this: the secular equation~(\ref{eq:secular})
gives a univocal correspondence (modulo trivial permutations) between
its coefficients and its solutions (that are physically acceptable in
the real domain).  Then, when all three $\xi_i$'s are different, we
can use $T,A,D$ instead of the $\xi_i$ to describe field space, and as
we are interested in a fully hierarchical pattern of Yukawas, we are
precisely in that situation.\footnote{In ref.~\cite{Nardi:2011st}, vev
  structures $\langle Y\rangle = v_Y\,{\rm
    diag}(\epsilon',\epsilon,1)$ were obtained by minimizing
  separately two functions respectively of $D$ and $A$, that included
  their logs, and assuming that the minimum of the two separate
  functions gave the global minimum.  That procedure is in fact
  incorrect. By minimizing numerically the full logarithmic potential
  we have verified that global minima with $\epsilon'=\epsilon$ are
  obtained.} In all other cases, when one of the three equations in
(\ref{simplemin}) is not satisfied, the vacuum will have at most the
partial hierarchy $\langle Y\rangle \sim {\rm diag}(a,a,b)$.

We can now understand from another point of view the negative result
found in the previous section in our search for a hierarchical minimum
by perturbing over the tree-level minimum $\langle Y\rangle^h=v_h\,
\,{\rm diag}(0,0,1)$. Writing the potential as ${V}=V_0+\Delta
V$, with the tree-level potential $V_0$ as in (\ref{eq:V3}) and
$\Delta V$ representing some small correction, either from loop
effects or higher order operators, the minimization equations
(\ref{simplemin}) become
\bear
\label{eq:DV}
2\lambda T - m^2 +\frac{\partial \Delta V}{\partial T}&=&0\ ,\nonumber\\
\lambda_A +\frac{\partial \Delta V}{\partial A}&=&0\ ,\nonumber\\
-2\mu +\frac{\partial \Delta V}{\partial D}&=&0\ .  
\eear 
Let us write $\langle T\rangle = \langle T\rangle_0 + \delta\langle
T\rangle$, where $\langle T\rangle_0$ is the value of $T$ at the tree
level minimum and $\delta\langle T\rangle$ a shift in this value due
to $\Delta V$, with analogous notations for $A$ and $D$.  The tree
level minimum corresponding to $\langle Y\rangle^h$ has $\langle
T\rangle_0\neq 0$ and $\langle A\rangle_0=\langle D\rangle_0= 0$.  Now
if one tries to analyze by means of equations (\ref{eq:DV}) how this
minimum can be perturbed by the correction $\Delta V$, the first key
observation is that the tree-level minimum does not satisfy the
tree-level form (i.e. with $\Delta V$ removed) of all these equations,
but only the first one, with $\langle T\rangle_0=m^2/(2\lambda)$. The
shift $\delta \langle T\rangle$ would then be obtained as \beq
2\lambda \delta \langle T\rangle = -\left.\frac{\partial \Delta
    V}{\partial T}\right|_0\ , \eeq with the subscript $0$ indicating
that the derivative is evaluated at the tree-level minimum. However,
the two other minimization equations are not consistent unless $\mu$
and $\lambda_A$ are suppressed to at least the order of the
perturbation $\Delta V$, perhaps being zero.  As already noticed in
section~\ref{sec:loop}, if $\mu$ and $\lambda_A$ are in fact zero,
there are 17 massless states at tree-level, and this clears up the
difficulties with Georgi-Pais' theorem~\cite{Georgi:1977hm}. In the
diagonal basis, the minimum condition reads $T=\xi_1+\xi_2+\xi_3=
m^2/(2\lambda)$ which is satisfied by points on a hyperspherical
surface of equivalent $\{\xi_i\}$ vevs, including hierarchical
ones. Which vev is eventually selected then would depend on radiative
corrections (or on other higher order effects). Nevertheless, the
one-loop potential does not resolve this degeneracy: when
$\mu=\lambda_A=0$, $\Delta V$ is a function of the $T$ invariant only,
{\it i.e.} no potential for $A$ and $D$ is generated at one-loop. This
can be checked explicitly using the results presented in the Appendix,
which calculates the contributions to the potential from scalar
self-interactions. In principle, interactions of the Yukawa fields
with other sectors of the theory could give additional contributions
that could change this picture. This, however, would require
introducing a significant model-dependence.

Let us examine next the structure of a potential tailored to give any
desired pattern of $R_{ii}$.  Constructing such generic potential is
in fact straightforward:\footnote{A potential of this form is given in
  appendix B of~\cite{Alonso:2011yg}.}
\begin{equation}
  \label{eq:tot}
  V\equiv  
\lambda\left(T- \langle T\rangle\right)^2+
\frac{1}{\Lambda^4}
\left(A-\langle A\rangle\right)^2
+
\frac{1}{\Lambda^2}
\left(D- \langle D\rangle\right)^2\ ,
\end{equation}
where the values of $\langle T\rangle, \langle A\rangle, \langle D\rangle$
correspond to the chosen values of $\xi_i$ according to 
\eqns{eq:T}{eq:D2}.
Omitting an unimportant constant, we can split $V=V_0+\Delta V$ 
in the usual renormalizable part
\begin{equation}
  \label{eq:ren}
  V_0=\lambda T^2-m^2 T + \lambda_A A- 2 \mu D\ ,
\end{equation}
plus the correction $\Delta V$, which contains the following two $d=8$
and $d=6$ terms:
\begin{equation}
  \label{eq:nonren2}
 \Delta  V= \frac{1}{\Lambda^4} A^2 + \frac{1}{\Lambda^2} D^2\, .
\end{equation}
For any (finite) cutoff scale $\Lambda$, any chosen vacuum
configuration $\{\langle T\rangle,\,\langle A\rangle,\, \langle
D\rangle\}$ can be reproduced by choosing the values of the parameters
as:
\beq 
\label{eq:vevs0}
m^2=2\lambda \langle T\rangle\ ,\;\;\; \lambda_A =
-2\frac{\langle A\rangle}{\Lambda^4} < 0\ , \;\;\; \mu = \frac{\langle
  D\rangle}{\Lambda^2}\ .  
\eeq 
For the validity of the effective theory approach, and to reproduce
the hierarchical Yukawas, the vacuum expectation values should be
smaller than the corresponding power of the scale $\Lambda$. This
means, in particular, that $\mu$ and $\lambda_A$ are required to be
quite suppressed also in this scenario.\footnote{For example, by
  adding to $V_A$, \eqn{eq:VA3}, higher order corrections in the form
  $(\lambda_A + \langle T\rangle/\Lambda^2 + \langle
  T\rangle^2/\Lambda^4 + \dots) A$, we would get from the hierarchy
  condition $\langle T\rangle^2 > \langle A\rangle$ and from the
  second relation in \eqn{eq:vevs0} that the tree level coupling
  $\lambda_A$ must be smaller than the second order correction $\sim
  \langle T\rangle^2/\Lambda^4$.}  Such suppression is directly
responsible for the Yukawa hierarchy, and this does not represent the
kind of natural explanation we are looking for.  Notice also that, as
$\lambda_A<0$ and $\mu>0$, the minimum of the tree-level potential
corresponds to the symmetric configuration $\xi_1=\xi_2=\xi_3$ so that
$\Delta V$ does not represent a small perturbation of the tree level
vacuum structure. The correction to the potential can have such a
large effect only because in this scenario the parameters of the
tree-level potential are assumed to have values much smaller than the
size of the corrections.

In the previous discussions, the scale dependence of the Lagrangian
parameters has been disregarded, although their RGE running can add
new features to the minimization problem.  For example, although the
functional form of the SM Higgs potential, $V(h)=-(m^2/2)
h^2+(\lambda_h/4) h^4$, allows for only one minimum, its
renormalization-improved version, with running $m^2$ and $\lambda_h$
evaluated at a renormalization scale $Q\sim h$ (as necessary for a
faithful description of the potential in a large range of $h$ values)
has a richer structure. In fact, for low values of the Higgs mass (as
the value currently measured at the LHC, $m_h\simeq 126$ GeV
\cite{mh}) a second minimum develops at a scale much larger than the
electroweak scale, when the quartic Higgs coupling $\lambda_h$ is
driven to negative values due to large radiative corrections from the
top Yukawa coupling (see ref.~\cite{Degrassi:2012ry} for the
state-of-the-art analysis).  For particular values of the Higgs mass,
the negative quartic stays small enough to cancel against the value of
its $\beta$ function, providing a second solution to the minimization
equation $dV/d h \simeq [\lambda_h + (1/4) d\lambda_h/d\log
h]h^3=0$. We could imagine something similar happening with the
potential for the Yukawa fields $Y$: the structure of this potential
at the renormalization scale relevant for the largest nonzero vev
$Q^2\sim \xi_3$ could in principle be different from its structure at
the lower scales relevant for the smaller vevs $\xi_1$ and
$\xi_2$. However, the problem is now complicated by the fact that the
potential is a multifield one and the correct description of a
hierarchical vacuum requires the use of three different
renormalization scales simultaneously (a multiscale problem often
faced in effective potential studies, see {\it e.g.}
\cite{multiscale}). We limit ourselves to pointing out this possibility
as worthy of future exploration.

\section{Symmetry breaking via  reducible representations}

The results of the previous sections make clear which way is left open
to get a phenomenologically viable pattern of vevs for the components
of the Yukawa field $Y$. Namely, the flavour symmetry must be broken
down to $H_{abc}$,~\eqn{eq:Hg}, already at the tree level. 
For this, we need a non-minimal set of scalar fields in reducible
representations of the flavour group. In fact, breaking a symmetry by
means of reducible representations avoids at once the issue of maximal
stability little groups: even defining what are {\it open strata} and
{\it closed strata} is not clear in this case.  A minimal enlargement
of the scalar sector involves adding two multiplets, $Z_{L,R}$, that
transform respectively in the fundamental of one of the two group
factors $SU(3)_L \times SU(3)_R$ while they are singlets under the
other one:\footnote{The vevs of $Z_{L,R}$ represent sources of flavour
  breaking that do not transform as the SM Yukawa couplings, and thus
  in principle imply a non Minimal Flavour Violating
  (MFV)~\cite{DAmbrosio:2002ex} scenario.  However, by a suitable
  choice of the representations of the messengers that generate the
  Yukawa operators, one can forbid all dangerous FCNC Yukawa-like
  operators involving $Z_{L,R}$.}
%
\begin{equation}
  \label{eq:ZLZR}
 Z_L = (\mathbf{3},\mathbf{1}), \qquad Z_R = (\mathbf{1},\mathbf{3})\,.
\end{equation}
The most general $SU(3)_L \times SU(3)_R$ invariant potential 
involving $Z_L$, $Z_R$ and $Y=(\mathbf{3},\mathbf{\bar 3})$  can be
written as
\begin{equation}
  \label{eq:VYZ}
  V = V_A+V_{\cal D} + V_l + V_m + V_{\tilde \nu}\,,  
\end{equation}
with
\begin{eqnarray}
\nonumber 
  V_l &=&
  \lambda\left(T-\frac{m^2}{2\lambda} \right)^2  +
  \lambda_L\left(|Z_L|^2-\frac{m^2_L}{2\lambda_L} \right)^2 +  
  \lambda_R\left(|Z_R|^2-\frac{m^2_R}{2\lambda_R} \right)^2   
  \\ 
  && \hspace{-0.8cm} + \, g 
\left[\left(T-\frac{m^2}{2\lambda}\right)+\frac{g_{1L}}{g}
\left(|Z_L|^2-\frac{m^2_L}{2\lambda_L} \right) +
 \frac{g_{1R}}{g}
    \left(|Z_R|^2- \frac{m^2_R}{2\lambda_R} \right) \right]^2\,,  
 \label{eq:Vlength}
\\ 
  V_m &=& 
  g_{2L}Z_{L}^{\dagger}YY^{\dagger}Z_{L}+g_{2R}
  Z_{R}^{\dagger}Y^{\dagger}YZ_{R}\,,
  \label{eq:Vmisalign}
\\ 
  V_{\tilde \nu} &=& \tilde \nu\, Z_{L}^{\dagger}YZ_{R} + {\rm h.c.}\,, 
  \label{eq:Vtnu}
  \end{eqnarray}
  while $V_A$ and $V_{\cal D}$ 
  have already been given in \eqn{eq:VA3} and \eqn{eq:VD3}. $V_l$ in
  \eqn{eq:Vlength} can be equivalently written as
\begin{eqnarray}
\nonumber
  \label{eq:Vl}
  V_l &=&  
  \hat\lambda \left(T - \frac{\hat m^2}{2\hat\lambda} \right)^2 
  +\hat \lambda_L \left( |Z_L|^2
  -\frac{\hat m_L^2}{2\hat\lambda_L}\right)^2  
  +\hat \lambda_R \left( |Z_R|^2
  -\frac{\hat m_R^2}{2\hat\lambda_R}\right)^2  \\
&+& 2 T\left(g_{1L} |Z_L|^2 +g_{1R} |Z_R|^2 \right)
+2\, \frac{g_{1L}\,g_{1R}}{g}\, |Z_L|^2\, |Z_R|^2 + \, {\rm const}\,. 
\end{eqnarray}
This second way of writing $V_l$ makes apparent which relevant terms
quadratic and quartic in the fields have been included; however, 
\eqn{eq:Vlength} is more convenient for minimization, since it makes
transparent that $V_l$ (which for $\lambda, \lambda_L, \lambda_R, g
>0$ has manifestly its minimum when it vanishes) determines the
`lengths' of the three multiplets to be, respectively:
\begin{equation}
  \label{eq:vevs}
\langle T\rangle\equiv v^2_Y =\frac{m^{2}}{2\lambda},\qquad 
\langle |Z_{L}|^{2}\rangle\equiv v^2_L =  \frac{m_{L}^{2}}{2\lambda_{L}},\qquad
\langle |Z_{R}|^{2}\rangle \equiv v^2_R = \frac{m_{R}^{2}}{2\lambda_{R}}
\,, 
\end{equation}
without having other effects on the particular structure and/or
alignment of the three vevs.  Note in particular that \eqn{eq:Vlength}
also makes apparent that the correct tree-level minima can already be
obtained from the first line alone, that is, by setting $g_{1L},
g_{1R}$ and $g \to 0$ (in particular, this limit largely simplifies the
identification of the Goldstone bosons).

The role of $V_m$ is instead that of misaligning (or aligning) the
vevs of $Z_{L,R}$ with the vev of $Y$. Our aim is to enforce a maximum
misalignment, in order to obtain the smallest possible intersection
among the little groups of the vevs of the three multiplets, and this
is achieved for $g_{2L},g_{2R}>0$. Thus we assume that both 
couplings in $V_m$ are positive, and we will not analyze other
possibilities, since they would only yield phenomenologically
uninteresting vacuum structures.

Once the matrix of constant fields $Y$ has been brought to the
diagonal and partially real form, \eqn{eq:basis}, by means of symmetry
rotations, we are left with the freedom of removing only one phase
from $Z_L$ and another one from $Z_R$. However, we will now argue
that, for the minimization problem, we can take $Z_{L,R}$ real.  After
defining $\nu = |\tilde \nu|$, the last term $V_{\tilde \nu}$ can be
rewritten as:
\begin{equation}
  \label{eq:tphinu}
 V_{\tilde \nu}
  = 2\,\nu\, \left|Z_{L}^{\dagger}YZ_{R}\right| \cos\phi_{LR}\,.
\end{equation}
At fixed values of $ \langle Y\rangle,\, \langle Z_L\rangle,\, \langle
Z_R\rangle$ the minimum of $V_{\tilde \nu}$ occurs when
$\cos\phi_{LR}=-1$, that is, for real and negative values of the vev
of the trilinear term $\langle Z_{L}^{\dagger}YZ_{R}\rangle$.  On the
other hand, it is easy to verify that the modulus 
of the trilinear term in \eqn{eq:tphinu}
is extremized for real values of the three vevs $ \langle Y\rangle,\, \langle Z_L\rangle,\, \langle
Z_R\rangle$.  
 We have already argued that minimization of the potential can be 
 carried out by taking $V_{\cal D}$ in the simplified form $V_D$ 
 given in \eqn{eq:VDY1}, which corresponds to real $Y$.   
 From \eqn{eq:tphinu} we can similarly conclude that, after including
$Z_{L,R}$, the minimization of the potential can be explored 
around the configuration
\begin{equation}
  \label{eq:phinu}
 V_{\tilde \nu}^{\rm min}\equiv V_\nu 
  = - 2\,\nu\, Z_{L}^{\dagger}YZ_{R}\,, 
\end{equation}
where $Z_{L,R}$ can also be taken to be real.  With this
simplification it is not difficult to work out the structure of the
vevs at the minimum of the potential.  Let us start by setting $\nu
\to 0$.  For $g_{2L},\,g_{2R}> 0 $ and $Y$ in diagonal form, $V_m$ is
always positive, and thus it is minimized when it vanishes, which
occurs when the vevs of $Z_{L,R}$ are misaligned with respect to the
vev of $Y$, as for example  $\langle Y\rangle = v_h\,{\rm
  diag}\left(0,0,1\right)$, $\langle Z_{L}\rangle =v_L\,
\left(c_{L},s_{L},0\right)$ and $\langle Z_{R}\rangle =v_R\,
\left(c_{R},s_{R},0\right)$, with $c_{L,R}^2+s_{L,R}^2=1$.
Thus, in this limit the minimum of the potential is $V^{\rm
  min}_{(\nu= 0)}=0$.  However, when $\nu\neq 0$ lower (negative)
values of $V_\nu$ become possible: for small but nonvanishing
values of the first two diagonal entries in $\langle Y\rangle$ the
negative sign of $V_\nu$ implies the possibility of adding a negative
contribution to the minimum of the potential. 
It is true that there is also a price to pay since 
the minimum of $V_m$ 
will then be lifted to positive values, but while this effect in $V_m$
is quadratic in the small $\langle Y\rangle$ entries, it is linear,
and thus dominant, in $V_\nu$.  We thus expect that $V_\nu$ can favour
configurations in which the zero entries in $\langle Y\rangle$ are
lifted to non-zero values.

To see explicitly how this can occur, let us take the new vacua
in the form: 
\begin{eqnarray}
  \label{eq:newvevs}
\nonumber
\langle Y\rangle = v_Y\,{\rm diag}\left(\epsilon,\epsilon',y\right), 
\quad &{\rm with}& \quad \epsilon,\epsilon'\ll y\,,
\qquad  \epsilon^2+{\epsilon'}^2+ y^2=1, \\
\nonumber
\langle Z_{L}\rangle = v_L\,
\left(z_{L},\epsilon'_{L},\epsilon_L\right)
\ \quad  &{\rm with}& \quad \epsilon'_{L},\epsilon_L\ll z_L, \quad 
{\epsilon'_{L}}^2+\epsilon_L^2+z_{L}^2=1, \\ 
\langle Z_{R}\rangle = v_R\,
\left(z_{R},\epsilon'_{R},\epsilon_R\right)
\quad &{\rm with}& \quad  \epsilon'_{R},\epsilon_R\ll z_R, \quad 
{\epsilon'_{R}}^2+\epsilon_R^2+z_{R}^2=1 \,.  
\end{eqnarray}
$V_l$ in \eqn{eq:Vlength} fixes the `lengths' $v_Y$ and $v_{L,R}$, and
vanishes.  So we need to consider only the effect of 
\begin{eqnarray}
  \label{eq:Veps1}
\nonumber
V_\epsilon &\equiv& V_A+V_D+V_m+V_\nu   \\
&=&  \lambda_A A-2\, \mu\, D +
g_{2L}Z_{L}^{\dagger}YY^{\dagger}Z_{L}+g_{2R}
Z_{R}^{\dagger}Y^{\dagger}YZ_{R}
-2\, \nu\, Z_{L}^{\dagger}YZ_{R}\,, 
\end{eqnarray}
which vanishes in the limit $\nu\to 0$ but remains of
${\cal O}(\epsilon)$ when   $\nu\neq 0$. 
Plugging into $V_\epsilon$ the vevs in \eqn{eq:newvevs} we obtain:
\begin{eqnarray}
\nonumber 
\frac{1}{v_Y^2}V_{\epsilon} 
& = & \lambda_A v_Y^2 \left[\epsilon^{2}\epsilon'^{2}+y^2\left(\epsilon'^{2}+
\epsilon^{2}\right)\right]-2\,v_Y\,\mu\,\epsilon\epsilon'y\nonumber \\
& + & 
g_{2L}\,v_L^2\, \left(z_{L}^{2}\epsilon^{2}+\epsilon_{L}'^{2}
\epsilon'^{2}+\epsilon_{L}^{2}\,y^{2}\right)
+g_{2R}\,v_R^2\, \left(z_{R}^{2}
\epsilon^{2}+\epsilon_{R}'^{2}\epsilon'^{2}+\epsilon_{R}^{2}y^2\right)
\nonumber \\
 & - &2\,\nu\,\frac{v_Lv_R}{v_Y}\,\left(z_{L}z_{R}\epsilon+\epsilon'_{L}
\epsilon'_{R}\epsilon'+\epsilon_{L}\epsilon_{R}\,y\right)\,.
\label{eq:V_epsilon}
\end{eqnarray}
The role of $V_\nu$, which is linear in $\epsilon$ in preferring
values of $\epsilon\neq 0$ is apparent, as well as the role of $V_D$
in favouring in turn $\epsilon'\neq 0$.  A simple illustrative solution
in terms of the relevant parameters $\mu$ and $\nu$ can be obtained by
setting for simplicity $v_{Y}=v_{L}=v_{R}$ and
$\lambda_A=g_{2L}=g_{2R}$ and by recalling that the lengths are fixed
(i.e. $y^2=1-\epsilon^2-{\epsilon'}^2$ etc.).  Solving for the
extremal conditions $\partial V_{\epsilon}/\partial\epsilon =\partial
V_{\epsilon}/\partial\epsilon' =\partial
V_{\epsilon}/\partial\epsilon_{L,R} =\partial
V_{\epsilon}/\partial\epsilon_{L,R}'=0$, and truncating to terms
$\mathcal{O}\left(\epsilon^{2}\right)$ we obtain a unique solution for
the global minimum:
\begin{equation}
\epsilon =  \frac{\lambda_A\, \nu\, v_Y}{3\lambda_A^2\,v_Y^2-\mu^{2}}\,, \qquad 
\epsilon' =  \frac{\mu}{\lambda_A\, v_Y}\, \epsilon\,, \qquad 
\epsilon_{L,R}  =  \epsilon_{L,R}'=0, 
\label{eq:finalvevs}
\end{equation}
with $V_\epsilon^{{\rm min}}=- \nu\, v_Y^3\,\epsilon$.  The relations in
\eqn{eq:finalvevs} show that phenomenologically acceptable vevs for
the up-quark Yukawa couplings could be obtained for example for $\nu
\sim \mu \sim 10^{-2}\cdot \lambda_A\, v_Y$, yielding $\epsilon' \sim
10^{-2}\cdot \epsilon \sim 10^{-4}$.  A second possibility, namely
$\nu \sim \lambda_A\, v_Y\sim 10^{-2}\cdot \mu$ yielding $\epsilon \sim
10^{-2}\cdot \epsilon' \sim 10^{-4}$ is in fact not viable since in
this case it is not possible to satisfy simultaneously the constraint
on $\mu/m$ from \eqn{eq:sol3} and the condition $v_Y\simeq
m/\sqrt{2\lambda}\sim \Lambda$ from the value of the top-quark Yukawa
coupling.

To verify the correctness of the simple minimization procedure that we
have outlined above, we have also performed a set of numerical
minimizations. This has confirmed the lifting of the zeroes in
$\langle Y\rangle$ to nonvanishing entries whose values, among other
things, are controlled in a crucial way by $\nu$ and $\mu$ that, in
order to reproduce the observed Yukawa hierarchies, should be somewhat
suppressed with respect to the other dimensional parameters
$m,\,m_L,\,m_R$. Let us stress however, that the numerical analysis
that we have carried out had just the scope of confirming the
structures obtained analytically, and that the extremum corresponds to
a real global minimum, and did not aim at a thorough exploration of
the full parameter space for finding sets of values satisfying
particular naturalness requirements.

One final interesting point is that in all cases we find that two
entries in both $\langle Z_{L,R}\rangle$ vanish, see
\eqn{eq:finalvevs}. The little group of the vevs of the two
fundamental representations is then $H_{LR}=SU(2)_L\times
SU(2)_R$. The intersection of $H_{LR}$ with the little group of the
vev of the bi-fundamental $H_Y=U(1)_{(\lambda_3)_{L+R}} \times
U(1)_{(\lambda_8)_{L+R}}$ is then  $ H_{LR} \cap H_Y =
U(1)_{(\sigma_3)_{L+R}}$, corresponding to the diagonal generator
$\sigma_3=\frac{1}{2}(\sqrt{3}\lambda_8-\lambda_3)$ of
$SU(2)_{L+R}$. This means that, out of the $8+8$ generators of the
flavour group, only one remains unbroken, and we can then predict
fifteen Goldstone bosons. We have studied numerically the spectrum of
scalar particles, and this confirmed that the symmetry breaking
pattern induced by scalars in the reducible representation
$(\mathbf{3},\mathbf{1})\oplus(\mathbf{1},\mathbf{3})\oplus
(\mathbf{3},\mathbf{\bar 3})$ is indeed $SU(3)_{L}\times SU(3)_{R} \to
U(1)_{(\sigma_3)_{L+R}}$.

\section{Conclusions}

The possibility that the highly non-symmetric spectra of the fermions
with the same SM quantum numbers could arise from the specific
structure of the vacuum of an otherwise flavour-symmetric theory, is
theoretically very attractive.  In this paper, we have studied which
premises are needed to render this idea phenomenologically viable.

In ref.~\cite{Nardi:2011st} it was found, as a promising starting
point, that an $SU(3)_L\times SU(3)_R$ invariant tree-level potential
for Yukawa fields $Y$ transforming as the irreducible bi-fundamental
$(\mathbf{3},\mathbf{\bar 3})$ representation of the flavour group
admits vacuum structures  $\langle Y\rangle \propto {\rm
  diag}(0,0,1)$. It was then natural to ask if the vanishing entries
could be lifted to suppressed but nonvanishing values by some type of
small perturbation. This would correspond to a stepwise breaking of the
flavour symmetry.

In this paper we have argued that the structure of the tree-level
vacua is perturbatively stable. Regarding the possibility of a
stepwise symmetry breaking due to loop corrections, it was already
discussed long ago~\cite{Georgi:1972mc,Georgi:1977hm} that this can
only occur in the presence of additional (non NGB) scalars that are
massless at tree level.  By direct computation of the effective
potential (see the Appendix) we have confirmed that the little groups
of the vacua of the tree-level potential are also the little groups of
the vacua of the one-loop corrected effective potential, and that this
remains true even in the presence of additional scalars that are {\it
  unnaturally} massless (that is, massless due to fine tuning of some
parameters).  We have also argued that this result can be
straightforwardly extended to the possible effects of operators of
higher dimension. In Section~\ref{sec:generalization} we have further
confirmed this and argued that, except for some special cases, a
generic $SU(3)_L\times SU(3)_R$ invariant function of a single scalar
field $Y$ transforming in the bi-fundamental representation of the
group, admits vacuum structures $\langle Y\rangle$ with at least two
equal eigenvalues, and thus it cannot yield a fully hierarchical
pattern.

We have thus learned that a phenomenologically viable Yukawa structure
$\langle Y\rangle \propto {\rm diag}(\epsilon,\epsilon',1)$ must arise
already at the tree level, and that in order to achieve this, the
breaking must occur via reducible representations.  After enlarging
the scalar sector by adding two multiplets $Z_L$ and $Z_R$
transforming respectively in the fundamental representation of the
$SU(3)_{L,R}$ factors of the flavour group, we have constructed the
most general fourth-order potential involving $Z_{L,R}$ and $Y$, and
we have shown that minima yielding the hierarchical structure $\langle
Y\rangle \propto {\rm diag}(\epsilon,\epsilon',1)$ can indeed appear.
The Yukawa hierarchy for the up-type quarks can then be qualitatively
reproduced at the cost of a relatively mild hierarchy between the
dimensional parameters of the scalar potential, not exceeding $\sim
10^{-2}$. The hierarchies in the down-quark and charged
lepton sectors are also reproducible with even milder hierarchies in
the fundamental parameters.

In a more complete scenario we would first need to extend the symmetry
to the full quark flavour group $SU(3)_{Q_L}\times SU(3)_{u_R}\times
SU(3)_{d_R}$, and then couple through appropriate renormalizable
invariant terms the Yukawa fields of the up and down-quark sectors
$Y^q,\, Z^q$ ($q=u,d$) and $Z^Q$ 
(see ref.~\cite{Nardi:2011st} for a
first attempt with only irreducible representations $Y^q$).  Besides
reproducing the mass hierarchies, such a scenario should also
reproduce the hierarchies in the CKM mixing angles, and yield a
nonvanishing value for the CP violating Jarlskog
invariant~\cite{Jarlskog:1985cw}.  We believe that, in spite of its
complexity, such a program can be carried out successfully, and we
expect to publish soon some results that go in this
direction~\cite{inpreparation}.

Eventually, one should also worry about the testability of this type
of construction. It is clear that the scalar
potential of a more complete model will contain a number of
fundamental parameters much larger than the number of observables (the
six quark masses, the three mixing angles, and the CP violating phase
$\delta$) implying that it is unlikely that predictive relations among
different observables could arise.  Direct evidences might arise from
the fact that if the flavour symmetry is global, then SSB implies the
presence of Nambu-Goldstone bosons that could show up in yet unseen
hadron decays or in rare flavour violating processes~\cite{Wilczek:1982rv}.  
If the flavour symmetry is instead gauged~\cite{Albrecht:2010xh,Grinstein:2010ve},
then to ensure the absence of gauge anomalies additional fermions must
be introduced~\cite{Grinstein:2010ve}, and their detection could
represent a smoking gun for this type of models. All this remains,
however, a bit speculative, especially because the theory provides no
hint of the scale at which the flavour symmetry gets broken,
and very large scales would suppress most, if not all, types of
signatures. In spite of these considerations, being able to reproduce
the observed pattern of Yukawa couplings from the SSB of the flavour
symmetry would certainly represent an important theoretical
achievement, and we believe that the results discussed here can 
provide some relevant steps in this direction.


\mathversion{bold}
\appendix
 \section{Effective potential for  
a single irreducible representation}
\mathversion{normal}

The $SU(3)_L\times SU(3)_R$ invariant one-loop Coleman-Weinberg
(CW)~\cite{Coleman:1973jx,Jackiw:1974cv} effective potential for a
scalar field $Y$ in the bi-fundamental representation
$(\mathbf{3},\mathbf{\overline{3}})$ can be written, in the
$\overline{\rm MS}$ scheme, as:

\begin{equation}
  \label{eq:effpot}
V = V_0+V_1
\end{equation}
with
\begin{eqnarray}
  \label{eq:effpot0}
V_0 &=& \lambda
\left[T(Y)-\frac{m^2}{2\lambda}\right]^2
+\lambda_A A(Y)+
\mu {\cal D}(Y) +
\mu^*{\cal D}^*(Y) 
  \\
\label{eq:effpot1}
V_1 &=& 
\frac{1}{64\pi^2}\, \sum_i 
M_i^4(Y)\left[\log\frac{M_i^2(Y)}{\Lambda^2}-\frac{3}{2}\right]
\end{eqnarray}
where all the parameters in $V_0$ are renormalized at the scale
$\Lambda$: $\lambda=\lambda(\Lambda),\,m^2=m^2(\Lambda),\,$etc.  The
field dependent mass functions $M_i^2(Y)$ in \eqn{eq:effpot1} are the
eigenvalues of the matrix
\begin{equation}
  \label{eq:M2ij}
[{\cal M}^2]_{ij,kl} = 
\frac{\partial^2 V_0}{\partial {\cal Y}_{ij} 
\partial {\cal Y}_{kl}}\,\bigg|_{Y} 
\end{equation}
where ${\cal Y}_{ij}=\{R_{ij},\,J_{ij}\}$ with $R_{ij}\,(J_{ij})=
\sqrt{2}\> {\rm Re}({\rm Im})Y_{ij} $ and, without loss of generality,
we can take the background constant field $Y$ in the diagonal form
\eqn{eq:basis}.  Here we are including only the contributions to the
effective potential that come from scalar self-interactions between
the components of the $Y$ field.  In concrete models $Y$ will interact
with other sectors of the theory, which will then also contribute 
to the loop corrected potential.

From~\eqn{eq:M2ij} one can compute straightforwardly the two traces
\begin{eqnarray}
\label{eq:M2} 
{\rm Tr}{\cal M}^{2} & = & 8\left(5\lambda+\lambda_A\right)T-18m^{2},\\
\label{eq:M4}
{\rm Tr}{\cal M}^{4} & = & 4T^{2}\left[(5\lambda+\lambda_A)^2
+(\lambda-\lambda_A)^{2}\right]-8T\left[2m^{2}\left(5\lambda+\lambda_A\right)-
\left|\mu\right|^{2}\right]\nonumber \\
 &  & +48\lambda\lambda_A A
+24\left(\lambda+\lambda_A\right)\left(\mu{\cal {\cal D}}
+\mu^{*}{\cal D}^{*}\right)+18m^{4}.  
\end{eqnarray}
which give the field dependent divergent part of the CW potential,
which (using a cutoff regularization) reads:
\begin{equation}
  \label{eq:div}
\delta_\Lambda V= 
\frac{\Lambda^2}{32\pi^2}\,{\rm Tr}{\cal M}^{2} + 
\frac{\log \Lambda^2}{64\pi^2}  \,  {\rm Tr}{\cal M}^{4}\,. 
\end{equation}
The logarithmic part can be used to obtain the beta functions of $\lambda$,
$\lambda_A$, $m^2$ and $\mu$ (up to contributions from other sectors
of the theory).

Computing the finite contribution to the CW potential \eqn{eq:effpot1}
is instead a difficult problem, since it requires diagonalizing the
full $18\times 18$ matrix \eqn{eq:M2ij}, namely solving the
eigenvalue equation
\begin{equation}
  \label{eq:A1}
 \det(M^2\cdot I -{\cal M}^2)=0\,,   
\end{equation}
where $M^2$ are the eigenvalues, $I \equiv I_{18\times 18}$ and ${\cal
  M}^2$ is given in \eqn{eq:M2ij} evaluated at $Y=\langle Y\rangle=
{\rm diag}( \sqrt{\xi_1}, \sqrt{\xi_2}, \sqrt{\xi_3})$.

The problem is somewhat simplified by the fact that the eigenvalues
will come in multiplets of the unbroken little group, and another
simplification is obtained by looking for a minimum around the
configuration \eqn{eq:VDY2}, that is $\langle Y\rangle = {\rm diag}(
\sqrt{\xi_1}, \sqrt{\xi_2}, \sqrt{\xi_3})$.\footnote{This is justified
  by the fact that after reabsorbing the phase of $\tilde\mu$ in
  ${\cal D}$, any imaginary part of the determinant, corresponding in
  our basis to $J_{33}\neq 0$, would imply deviation from
  $\cos\phi_{\cal D}=-1$ and would shift the value of $V_{\cal
    D}$~\eqn{eq:VD3} away from its minimum.  In the loop corrections,
  functions involving terms like ${\cal D}^n+{\cal D^*}^n\sim D^n
  \cos(n\,\phi_{\cal D})$ also appear.  For even $n$ they are
  extremized at $\phi_{\cal D}=\pi/n$ and could thus shift, at least
  in principle, the minimum from $\pi$ to the doubly degenerated point
  $\phi_{\cal D}=\pi \pm \alpha$ (with suitable values of $\alpha$),
  yielding a complex $\langle {\cal D}\rangle$ and spontaneous
  breaking of CP.  However, the loop coefficient suppressing these
  contributions is small enough to guarantee stability of the minimum
  in $\pi$ and that the background value of the determinant does not
  acquire an imaginary part. This is of course in agreement with
  ref.~\cite{Georgi:1977hm} where the necessary conditions for stepwise
  breaking of discrete symmetries were stated.}  We have managed to
solve the problem by means of a `brute force' procedure, that can be
resumed in the following steps:

(i) The determinant in \eqn{eq:A1} admits a factorization of the form
\begin{equation}
  \label{eq:factor}
  \det(M^2\cdot I -{\cal M}^2) = P^{(6)}(M^2)\times 
\Pi_{i=1}^3 \Pi_\pm 
  (M^2 - M^2_{i\pm})^2 \,, 
\end{equation}
where $P^{(6)}(M^2)$ is a sixth-order polynomial in $M^2$.  This
factorization allows to identify the first twelve eigenvalues (labeled
with $i\pm$ with $i=1,2,3)$ that come arranged into six degenerate
doublets, and are:
\beq 
M^2_{i\pm} = \overline{m}^{2}+\frac{1}{2}\lambda_A\, \xi_i
\pm\sqrt{F\left(\xi_i,T,A,D\right)},
\label{eq:deg_eigenvalues}
\eeq 
where
\beqa
\overline{m}^{2} & \equiv & -m^{2}+2\lambda T,\\
F\left(\xi_{i},T,A,D\right) & = &
\frac{1}{2}\left[{\lambda_A}^{2}\left(\frac{\xi_i}{2}-T\right
  )+\mu^{2}\right]\xi_i+\lambda_A\left(\lambda_A A+2\mu D\right). 
\eeqa
The eigenvalues $\xi_i$ of the matrix of constant classical fields
$Y Y^\dagger$, and can be explicitly written in terms of the invariants
$T,\,A,\,D$ by solving the cubic equation:
\beq 
\det(\xi I_{3\times 3}-Y Y^\dagger)
= 
\xi^{3}- T\xi^{2} + A\xi-D^{2} = 0.
\eeq 

(ii) Concerning the roots of  $P^{(6)}(M^2)$ in
\eqn{eq:factor}, it is well known that there is no formula in radicals
to solve polynomial equations beyond
quartic~\cite{Ruffini-Abel}. However, on physics grounds we know that
$P^{(6)}(M^2)=0$ must be solvable with real and positive solutions.
We have then approached the problem of extracting the solutions by
studying various limits with increasing steps of complexity:
$\lambda_A,\mu = 0$ ($\lambda\neq 0$); $\lambda,\mu = 0$ ($\lambda_A\neq
0$); $\mu = 0$ and $\lambda,\,\lambda_A\neq 0$; $\lambda_A=0$ and
$\lambda,\,\mu\neq 0$. This allowed us to identify some characteristic
structures appearing in the solutions.

(iii) Finally, given that $P^{(6)}(M^2)=0$ is solvable, it follows
that the sixth order polynomial must be factorizable in several
different ways into lower order polynomials like
$P^{(3)}\cdot P^{(3)} $ or $P^{(2)}\cdot P^{(2)} \cdot P^{(2)} $ or
$P^{(2)}\cdot P^{(4)}$,  which all must have real and positive
solutions.  This implies that the structure of the solutions can be
recast in the general form of solutions of quadratic,  cubic
and quartic polynomial equations.
When written down as the roots of two cubic polynomials the remaining six
eigenvalues read (with $i,i'=\{1,2,3\})$:
\beqa 
M^2_{i} & = & \overline{m}^{2}+\frac{2}{3}\lambda_A T
+\frac{r_i}{3}P^{1/3}+\frac{1}{3r_i}\frac{Z_P}{P^{1/3}}\ ,\nonumber \\
M^2_{i'} & = & \overline{m}^{2}+\frac{2}{3}\lambda_AT+\frac{4}{3}\lambda T
+\frac{r_{i'}}{3}Q^{1/3}+\frac{1}{3r_{i'}}\frac{Z_Q}{Q^{1/3}}\ ,
\label{eq:dis_eigenvalues}
\eeqa 
where $r_{i,i'}$ are the three roots of $x^3+1=0$, that is, $\{-1,e^{-i\pi/3},e^{i\pi/3}\}$, while 
\beqa 
Z_P & = & \lambda_A^{2}\left(T^{2}-3A\right)+3\mu^{2}T,\nonumber \\
Z_Q & = & Z_P+ 16\lambda^{2}T^{2}+4\lambda_A
\left[2\lambda\left(9A-T^{2}\right)
+3\lambda_A A\right]
+36\left(2\lambda+\lambda_A\right)\mu D,
\eeqa 
and 
\beqa 
P & = & \frac{1}{2}\sqrt{X_{P}^{2}-4Z_{P}^{3}}
-9\lambda_A\mu^{2}T^{2}-\frac{9}{2}\lambda_A^{3}AT
+\lambda_A^{3}T^{3}\nonumber \\
 &  & +\frac{27}{2}\left[2\mu^{3}D+2\lambda_A\mu^{2}A
+\lambda_A^{3}D^{2}\right],\nonumber \\
Q & = & -\frac{1}{2}\sqrt{X_{Q}^{2}-4Z_{Q}^{3}}
+9\left(4\lambda-\lambda_A\right)\mu^{2}T^{2}
-\left(4\lambda-\lambda_A\right)^{3}T^{3}\nonumber \\
 &  & -\frac{27}{2}\left[32\lambda^{2}\mu D
+32\lambda\lambda_A\left(\lambda A+\mu D\right)
-\lambda_A\left(4\lambda A-8\mu D+\lambda_AA\right)\right]T\nonumber \\
 &  & -\frac{27}{2}\left[2\mu^{3}D+16\lambda\mu^{2}A+2\lambda_A\mu^{2}A
+27\lambda_A^{2}\left(4\lambda+\lambda_A\right)D^{2}\right],
\eeqa 
with
\beqa 
X_{P} & = & 54\mu^{3}D-18\lambda_A\mu^{2}\left(T^{2}-3A\right)
+\lambda_A^{3}\left(2T^{3}-9AT+27D^{2}\right),\nonumber \\
X_{Q} & = & 2\left[64\lambda^{3}T^{3}-36\lambda\mu^{2}T^{2}
+432\lambda^{2}\mu DT+27\mu^{2}\left(8\lambda A+\mu D\right)\right]\nonumber \\
 &  & +6\lambda_A\left[-16\lambda^{2}T^{3}+3\mu^{2}T^{2}
+144\lambda\mu DT+9A\left(16\lambda^{2}T+\mu^{2}\right)\right]\nonumber \\
 &  & +3\lambda_A^{2}\left(-36\lambda AT+8\lambda T^{3}
+972\lambda D^{2}+72\mu DT\right)\nonumber \\
 &  & -\lambda_A^{3}\left(27AT+2T^{3}-729D^{2}\right)\ .
\eeqa 
By solving the eigenvalue problem numerically, and confronting
the solutions with the corresponding numerical values of the analytic
expressions for $M^2_i$, we have verified the correctness 
of the analytic   formulas given in~\eqns{eq:deg_eigenvalues}{eq:dis_eigenvalues}.

This explicit expression for the potential can then be used to explore
the change of the tree-level vacuum induced by the radiative
corrections. An alternative approach (see {\it e.g.} \cite{CPR}) that
does not require solving for the exact mass eigenvalues that enter the
CW potential is to calculate the field derivatives of
the potential (or derivatives with respect to the invariants $T,A,D$)
needed in the minimization equations using perturbation theory around
the tree-level vacuum, obtaining the derivatives of the masses
directly from the derivatives of the characteristic polynomial 
$\det(M^2\cdot I-{\cal M}^2)=0$ (which can be written in terms of the
$T,A,D$ invariants, although we do not write those expressions
explicitly). We have also verified in this manner that the one-loop
corrections shift the tree-level minimum rotating it in field space,
but without changing its minimum structure, which still yields 
a Yukawa coupling matrix of the form
$\langle Y \rangle \propto {\rm diag} (0,0,1)$.

\subsection*{Acknowledgments}
J.R.E. thanks the theory dep. of INFN at Padova Univ. for hospitality
while this work was in progress.  This work has been partly supported
by Spanish Consolider Ingenio 2010 Programme CPAN (CSD2007-00042) and
the Spanish Ministry MICNN under grants FPA2010-17747 and
FPA2011-25948; and the Generalitat de Catalunya grant 2009SGR894.



\begin{thebibliography}{99}

\bibitem{Froggatt:1978nt}
  C.~D.~Froggatt, H.~B.~Nielsen,
  {\it Hierarchy of Quark Masses, Cabibbo Angles and CP Violation}, 
  Nucl.\ Phys.\  {\bf B147}, 277 (1979).


 \bibitem{MFV}
 R.~S.~Chivukula, H.~Georgi,
  {\it Composite Technicolor Standard Model},
   Phys.\ Lett.\  {\bf B188}, 99 (1987). 


\bibitem{Anselm:1996jm} 
  A.~Anselm and Z.~Berezhiani,
  {\it Weak mixing angles as dynamical degrees of freedom,}
  Nucl.\ Phys.\ B {\bf 484}, 97 (1997)
  [hep-ph/9605400].

\bibitem{Berezhiani:2001mh} 
  Z.~Berezhiani and A.~Rossi,
  {\it Flavor structure, flavor symmetry and supersymmetry,}
  Nucl.\ Phys.\ Proc.\ Suppl.\  {\bf 101}, 410 (2001)
  [hep-ph/0107054].

\bibitem{Koide:2008qm}
  Y.~Koide,
{\it Phenomenological Meaning of a Neutrino Mass Matrix Related to 
 Up-Quark Masses},
  Phys.\ Rev.\  {\bf D78}, 093006 (2008) 
  [arXiv:0809.2449].

\bibitem{Koide:2008tr}
  Y.~Koide,
{\it Charged Lepton Mass Relations in a Supersymmetric Yukawaon Model},
  Phys.\ Rev.\  {\bf D79}, 033009 (2009) 
  [arXiv:0811.3470].

\bibitem{Koide:2012fw} 
  Y.~Koide and H.~Nishiura,
 {\it Yukawaon Model with U(3)$\times$S$_3$ Family Symmetries},
  arXiv:1202.5815.

\bibitem{Feldmann:2009dc}
  T.~Feldmann, M.~Jung, T.~Mannel,
  {\it Sequential Flavour Symmetry Breaking},
  Phys.\ Rev.\  {\bf D80}, 033003 (2009) 
  [arXiv:0906.1523].

\bibitem{Albrecht:2010xh}
 M.~E.~Albrecht, T.~.Feldmann, T.~Mannel,
{\it Goldstone Bosons in Effective Theories with Spontaneously 
Broken Flavour Symmetry}, 
 JHEP {\bf 1010 } (2010)  089 [arXiv:1002.4798]. 


\bibitem{Grinstein:2010ve} B.~Grinstein, M.~Redi and G.~Villadoro,
  {\it Low Scale Flavor Gauge Symmetries}, JHEP {\bf 1011}, 067 (2010)
  [arXiv:1009.2049].

\bibitem{Alonso:2011yg}
  R.~Alonso, M.~B.~Gavela, L.~Merlo, S.~Rigolin,
  {\it On The Potential of Minimal Flavour Violation}, 
   JHEP {\bf 1107}, 012 (2011)   [arXiv:1103.2915].

\bibitem{Nardi:2011st} 
  E.~Nardi,
{\it Naturally large Yukawa hierarchies}, 
  Phys.\ Rev.\ D {\bf 84}, 036008 (2011)
  [arXiv:1105.1770].

\bibitem{Mohapatra:2012km} 
  R.~N.~Mohapatra,
  {\it Gauged Flavor, Supersymmetry and Grand Unification,}
  arXiv:1205.6190  .

\bibitem{Xing:2007fb} 
  Z.~-z.~Xing, H.~Zhang and S.~Zhou,
   {\it Updated Values of Running Quark and Lepton Masses,}
  Phys.\ Rev.\ D {\bf 77}, 113016 (2008)
  [arXiv:0712.1419].


\bibitem{Michel:1971th} 
  L.~Michel and L.~A.~Radicati,
   {\it Properties of the breaking of hadronic internal symmetry,}
  Annals Phys.\  {\bf 66}, 758 (1971). See also 
  R.~Slansky,
   {\it Group Theory for Unified Model Building,}
  Phys.\ Rept.\  {\bf 79}, 1 (1981), Sect. 9.


\bibitem{JSKim}
  J.~Kim,
 {\it General Method For Analyzing Higgs Potentials,}
  Nucl.\ Phys.\ B {\bf 196} (1982) 285.


\bibitem{Michel-conj} 
L. Michel, CERN-TH-2716 (1979), contribution to
  the A. Visconti seminar.


\bibitem{Michel-counterex} 
  S.~Meljanac,
   {\it Origin Of Counter Examples To Michel's Conjecture,}
  Phys.\ Lett.\ B {\bf 168}, 371 (1986);
%
  M.~Abud, G.~Anastaze, P.~Eckert and H.~Ruegg,
   {\it Counter Example To Michel's Conjecture,}
  Phys.\ Lett.\ B {\bf 142}, 371 (1984);
%
   {\it Minima Of Higgs Potentials Corresponding To Nonmaximal Isotropy Subgroups,}
  Annals Phys.\  {\bf 162}, 155 (1985); 
%
  C.~J.~Cummins and R.~C.~King,
   {\it Absolute Minima Of The Higgs Potential For The 75 Of Su(5),}
  J.\ Phys.\ A A {\bf 19}, 161 (1986);
%
  T.~Hubsch, S.~Meljanac, S.~Pallua and G.~G.~Ross,
   {\it The Missing Multiplet Mechanism And 75 Breaking Of Supersymmetric Su(5),}
  Phys.\ Lett.\ B {\bf 161}, 122 (1985).




\bibitem{Georgi:1972mc} 
  H.~Georgi and S.~L.~Glashow,
   {\it Spontaneously broken gauge symmetry and elementary particle masses,}
  Phys.\ Rev.\ D {\bf 6}, 2977 (1972).


\bibitem{Georgi:1977hm} 
  H.~Georgi and A.~Pais,
   {\it Natural Stepwise Breaking of Gauge and Discrete Symmetries,}
  Phys.\ Rev.\ D {\bf 16}, 3520 (1977).


\bibitem{Coleman:1973jx} 
  S.~R.~Coleman and E.~J.~Weinberg,
   {\it Radiative Corrections as the Origin of Spontaneous Symmetry Breaking,}
  Phys.\ Rev.\ D {\bf 7}, 1888 (1973).

\bibitem{Georgi:1975tz} 
  H.~Georgi and A.~Pais,
   {\it Vacuum Symmetry and the PseudoGoldstone Phenomenon,}
  Phys.\ Rev.\ D {\bf 12}, 508 (1975).


\bibitem{multiscale}
  M.~B.~Einhorn and D.~R.~T.~Jones,
   {\it A New Renormalization Group Approach To Multiscale Problems,}
  Nucl.\ Phys.\ B {\bf 230} (1984) 261;
  M.~Bando, T.~Kugo, N.~Maekawa and H.~Nakano,
  {\it Improving the effective potential: Multimass scale case,}
  Prog.\ Theor.\ Phys.\  {\bf 90} (1993) 405
  [hep-ph/9210229];
  C.~Ford and C.~Wiesendanger,
   {\it A Multiscale subtraction scheme and partial renormalization group equations in the O(N) symmetric phi**4 theory,}
  Phys.\ Rev.\ D {\bf 55} (1997) 2202
  [hep-ph/9604392];
  C.~Ford and C.~Wiesendanger,
   {\it Multiscale renormalization,}
  Phys.\ Lett.\ B {\bf 398} (1997) 342
  [hep-th/9612193];
  J.~A.~Casas, V.~Di Clemente and M.~Quiros,
   {\it The Effective potential in the presence of several mass scales,}
  Nucl.\ Phys.\ B {\bf 553} (1999) 511
  [hep-ph/9809275].

\bibitem{mh}
  G.~Aad {\it et al.}  [ATLAS Collaboration],
 {\it Observation of a new particle in the search for the Standard Model Higgs boson with the ATLAS detector at the LHC,}
  Phys.\ Lett.\ B {\bf 716} (2012) 1
  [arXiv:1207.7214 [hep-ex]];
  S.~Chatrchyan {\it et al.}  [CMS Collaboration],
  {\it Observation of a new boson at a mass of 125 GeV with the CMS experiment at the LHC,}
  Phys.\ Lett.\ B {\bf 716} (2012) 30
  [arXiv:1207.7235 [hep-ex]].

\bibitem{Degrassi:2012ry} 
  G.~Degrassi, S.~Di Vita, J.~Elias-Miro, J.~R.~Espinosa, G.~F.~Giudice, G.~Isidori and A.~Strumia,
   {\it Higgs mass and vacuum stability in the Standard Model at NNLO,}
  JHEP {\bf 1208}, 098 (2012)
  [arXiv:1205.6497].


 \bibitem{DAmbrosio:2002ex}
  G.~D'Ambrosio, G.~F.~Giudice, G.~Isidori and A.~Strumia,
  {\it Minimal flavour violation: An effective field theory approach},
  Nucl.\ Phys.\  B {\bf 645}, 155 (2002)
  [arXiv:hep-ph/0207036].


\bibitem{Jarlskog:1985cw} 
  C.~Jarlskog,
   {\it A Basis Independent Formulation of the Connection Between Quark Mass Matrices, CP Violation and Experiment,}
  Z.\ Phys.\ C {\bf 29}, 491 (1985).


\bibitem{inpreparation}
C.~S.~Fong and E.~Nardi, in preparation. 

\bibitem{Wilczek:1982rv}
  F.~Wilczek,
  Phys.\ Rev.\ Lett.\  {\bf 49} (1982) 1549.

\bibitem{Jackiw:1974cv} 
  R.~Jackiw,
   {\it Functional evaluation of the effective potential,}
  Phys.\ Rev.\ D {\bf 9}, 1686 (1974).

\bibitem{Ruffini-Abel} P. Ruffini, ``Teoria generale delle Equazioni,
  in cui si dimostra impossibile la soluzione algebraica delle
  equazioni generali di grado superiore al quarto.''  Bologna,
  Stamperia di S. Tommaso d' Aquino, (1799). In Opere Matematiche di
  Paolo Ruffini a cura di E. Bortolotti, vol. 1, pp. 1-324, Cremonese,
  Roma, (1953); N. H. Abel, ``Beweis der Unm\"oglichkeit, algebraische
  Gleichungen von h\"oheren Graden als dem vierten allgemein
  aufzul\"osen,'' J. reine angew. Math. 1, 65, (1826). Reprinted in
  Abel, N. H.(Ed. L. Sylow and S. Lie). Christiania [Oslo], Norway,
  (1881). Reprinted in New York: Johnson Reprint Corp., pp. 66-87,
  (1988);

\bibitem{CPR}
  D.~Comelli, M.~Pietroni and A.~Riotto,
   {\it On the spontaneous CP breaking at finite temperature in a nonminimal supersymmetric model,}
  Phys.\ Rev.\ D {\bf 50} (1994) 7703
  [hep-ph/9406368].


\end{thebibliography}
   \end{document}